\documentclass[12pt]{article}

\newcommand{\be}{\begin{equation}} \newcommand{\ee}{\end{equation}}
\newcommand{\ba}{\begin{eqnarray}} \newcommand{\ea}{\end{eqnarray}}
\newcommand{\bit}{\begin{itemize}} \newcommand{\eit}{\end{itemize}}
\newcommand{\ben}{\begin{enumerate}} \newcommand{\een}{\end{enumerate}}

 \newcommand{\dpa}{\partial}

\begin{document}
%\large

\hspace*{0.29\textwidth} 
 \underline{\bf http://xxx.lanl.gov/e-print/physics/0111014}\\

\begin{center}
%%%%%%%%%%%%%%%%%%%%%%%%%%%%%%%%%%%%%%%%%%%%%%%%%%%%%%%%%%%%%%%%%%
%\title{%
\textbf{\Large
Damping of transversal plasma-electron oscillations and waves\\ 
       in low-collision electron-ion  plasmas}
%%%%%%%%%%%%%%%%%%%%%%%%%%%%%%%%%%%%%%%%%%%%%%%%%%%%%%%%%%%%%%%%%%
%\author{%
\vspace{3mm}

V.~N.~Soshnikov
%%%%%%%%%%%%%%%%%%%%%%%%%%%%%%%%%%%%%%%%%%%%%%%%%%%%%%%%%%%%%%%%%%
\footnote{Krasnodarskaya str., 51-2-168, Moscow 109559, Russia.}
%}
%%%%%%%%%%%%%%%%%%%%%%%%%%%%%%%%%%%%%%%%%%%%%%%%%%%%%%%%%%%%%%%%%%
%\address{%
\vspace{1mm}

Plasma Physics Dept.,\\
All-Russian Institute of Scientific and Technical Information\\
of the Russian Academy of Sciences\\
(VINITI, Usievitcha 20, 125315 Moscow, Russia)
%}%
%%%%%%%%%%%%%%%%%%%%%%%%%%%%%%%%%%%%%%%%%%%%%%%%%%%%%%%%%%%%%%%%%%
%\maketitle
\end{center}
\vspace{-2mm}

\begin{abstract}
  Previously developed method~\cite{bib-1,bib-3} for finding asymptotic solutions 
  of Vlasov equations using two-dimensional 
  (in coordinate $x$ and time $t$) Laplace transform
  is here applied to consider transversal oscillations and waves
  in low-collision quasi-neutral ($n_i \simeq n_e$)
  Maxwellian electron-ion  plasmas.
  We obtain two branches of electron waves:
  the ubiquitous one of  high-frequency and high-velocity oscillations 
  and the unusual low-velocity one.
  Taking into account Coulomb collisions in the limit
  $m_e \ll m_i$, $\bar{v_i} \ll \bar{v_e}$, and $T_e m_e \ll T_i m_i$
  results in expressions for transversal plasma-electron 
  oscillation/wave decrements with a damping 
  of the low-velocity electron branch $\sim n_i^{1/3}/\bar{v}_e^{4/3}$,
  where $n_i$ is the ion density 
  and $\bar{v}_e$ is the mean electron velocity. 
  It ought to rehabilitate Vlasov principal value prescription 
  for relevant integrals,
  but to supplement it with representation of an asymptotical solution 
  as a sum of exponents (not a single one!).
  "Non-damping" kinematical waves in low-collision plasma
  transform in the damping ones 
  at reasonably chosen iteration process.
\end{abstract}



%%%%%%%%%%%%%%%%%%%%%%%%%%%%%%%%%%%%%%%%%%%%%%%%%%%%%%%%%%%%%%%%%%
\section{Introduction}
%%%%%%%%%%%%%%%%%%%%%%%%%%%%%%%%%%%%%%%%%%%%%%%%%%%%%%%%%%%%%%%%%%

  Propagation of plane transversal electromagnetic waves in plasmas
is described by asymptotic in coordinates and time solutions
of the coupled system of kinetic equations for electrons and ions
and Maxwell equations for the electric field:
\ba
 \label{eq-1}
 \frac{\dpa f^{(e)}_1}{\dpa t}
  + v_x \frac{\dpa f^{(e)}_1}{\dpa x}
  - \frac{|e|E_z(x,t)}{m_e}\frac{\dpa f^{(e)}_0}{\dpa v_z}
  &=& 0\ ,\qquad\qquad \\ \nonumber && \\
 \label{eq-2}
 \frac{\dpa f^{(i)}_1}{\dpa t}
  + v_x \frac{\dpa f^{(i)}_1}{\dpa x}
  + \frac{|e|E_z(x,t)}{m_i}\frac{\dpa f^{(i)}_0}{\dpa v_z}
  &=& 0\ ,\qquad\qquad
\ea
\be
 \label{eq-3}
 \frac{\dpa^2 E_z(x,t)}{\dpa x^2}
  - \frac{1}{c^2} \frac{\dpa^2 E_z(x,t)}{\dpa t^2}
  + \frac{4\pi|e|}{c^2}\frac{\dpa }{\dpa t} \int v_z 
         \left(
          n_e f^{(e)}_1 - n_i f^{(i)}_1
         \right) d\vec{v}
 \ =\ 0\ ,
\ee
where
\be\label{eq-4}
 \int f^{(e,i)}_0 d\vec{v} = 1\ ;\quad 
 \left|f^{(e,i)}_1\right| \ll f^{(e,i)}_0\ ;\quad
 \left|\int v_z f^{(i)}_1 d\vec{v}\ \right| 
  \ll \left|\int v_z f^{(e)}_1 d\vec{v}\ \right|\ ,
\ee
the plane wave is moving in $x$-direction, 
$f_0$ is Maxwell distribution function.
Let us note that one can add to the right hand sides of
Eqs.(\ref{eq-1}), (\ref{eq-2}) some collision integrals.

Here we limit ourselves to a consideration of the particular case 
of plane electron waves 
in infinite homogenous fully ionized plasma
with the boundary electric field perturbation
$E(0,t)=E_0\exp(i\omega t)$.

In close analogy to our previous paper~\cite{bib-1},
where we have considered the damping of longitudinal waves
in the electron-ion low-collision plasmas,
the problem is solved using iteration technique
with substitution of collisionless distribution functions
into the Coulomb collision integrals.

%%%%%%%%%%%%%%%%%%%%%%%%%%%%%%%%%%%%%%%%%%%%%%%%%%%%%%%%%%%%%%%%%%
\section{The null iteration}
%%%%%%%%%%%%%%%%%%%%%%%%%%%%%%%%%%%%%%%%%%%%%%%%%%%%%%%%%%%%%%%%%%

   Neglecting the ion constituent in Eq.(\ref{eq-3})
and using Laplace transforms in Eqs.(\ref{eq-1})--(\ref{eq-3})
we arrive at
\ba
 E_z\left(x,t\right)
 &=&\frac{1}{\left(2\pi i\right)^2}
     \int\limits_{\sigma_1^-}^{\sigma_1^+}\!\!
      \int\limits_{\sigma_2^-}^{\sigma_2^+}
       E_{p_1p_2} e^{p_1 t+p_2 x} dp_1 dp_2 \ ,
\label{eq-5}\\
 f^{(e)}_1\left(\vec{v},x,t\right)
 &=&\frac{1}{\left(2\pi i\right)^2}
     \int\limits_{\sigma_1^-}^{\sigma_1^+}\!\!
      \int\limits_{\sigma_2^-}^{\sigma_2^+}
       f^{(1)}_{p_1p_2} e^{p_1 t+p_2 x} dp_1 dp_2 \ ,
\label{eq-6}\\
 \frac{\dpa f^{(e)}_1\left(\vec{v},x,t\right)}{\dpa x}
 &=&\ \frac{1}{\left(2\pi i\right)^2}
     \int\limits_{\sigma_1^-}^{\sigma_1^+}\!\!
      \int\limits_{\sigma_2^-}^{\sigma_2^+}
       \left(p_2 f^{(1)}_{p_1p_2}-f^{(1)}_{p_1}\right) 
        e^{p_1 t+p_2 x} dp_1 dp_2\ ,
\label{eq-7}\\
 \frac{\dpa f^{(e)}_1\left(\vec{v},x,t\right)}{\dpa t}
 &=&\ \frac{1}{\left(2\pi i\right)^2}
     \int\limits_{\sigma_1^-}^{\sigma_1^+}\!\!
      \int\limits_{\sigma_2^-}^{\sigma_2^+}
       \left(p_1 f^{(1)}_{p_1p_2}-f^{(1)}_{p_2}\right) 
        e^{p_1 t+p_2 x} dp_1 dp_2\ ,
\label{eq-8}\\
 \frac{\dpa^2 E_z(x,t)}{\dpa x^2}
 &=&\ \frac{1}{\left(2\pi i\right)^2}
     \int\limits_{\sigma_1^-}^{\sigma_1^+}\!\!
      \int\limits_{\sigma_2^-}^{\sigma_2^+}
       \left(p_2^2 E_{p_1p_2}-p_2E_{p_1}-F_{p_1}\right) 
        e^{p_1 t+p_2 x} dp_1 dp_2\ ,
\label{eq-9}\\
 \frac{\dpa^2 E_z(x,t)}{\dpa t^2}
 &=&\ \frac{1}{\left(2\pi i\right)^2}
     \int\limits_{\sigma_1^-}^{\sigma_1^+}\!\!
      \int\limits_{\sigma_2^-}^{\sigma_2^+}
       \left(p_1^2 E_{p_1p_2}-p_1E_{p_2}-F_{p_2}\right) 
        e^{p_1 t+p_2 x} dp_1 dp_2\ ,
\label{eq-10}
\ea
where $\sigma_{1,2}^{\pm}\equiv\sigma_{1,2}\pm i\infty$ and
$ f^{(1)}_{p_1}$, $f^{(1)}_{p_2}$, 
$F_{p_1}$, $F_{p_2}$, $E_{p_1}$, and 
$E_{p_2}$ are, correspondingly, 
Laplace transforms of
$f^{(e)}\left(\vec{v},0,t\right)$, $f^{(e)}\left(\vec{v},x,0\right)$,
$\frac{\dpa E_z(x,t)}{\dpa x}\Big|_{x=0}$,
$\frac{\dpa E_z(x,t)}{\dpa t}\Big|_{t=0}$,
$E_z(0,t)$, and $E_z(x,0)$.

Neglecting for simplicity initial and boundary values 
\be \label{eq-11}
 f^{(1)}_{p_1}\ ,\quad 
 f^{(1)}_{p_2}\ ,\quad
 F_{p_1}\ ,\quad
 F_{p_2}\ ,\quad
 E_{p_2}
\ee
(they do not affect characteristic frequencies $\omega \equiv -ip_1$
 and wave numbers $k \equiv -ip_2$)
one obtains the following equation for the double poles in $p_1$ and $p_2$:
\be
  E_{p_1p_2}
   \left[
     p_2^2 - \frac{p_1^2}{c^2}
   + \frac{\omega_L^2p_1}{c^2}\int 
      v_z \frac{\dpa f^{(e)}_{0}}{\dpa v_z}
           \frac{d^3\vec{v}}{p_1+v_xp_2} 
   \right]
= p_2 E_{p_1}\ ,
\label{eq-12}
\ee
where $\omega_L\equiv\sqrt{4\pi e^2n_e/m_e}$ 
is Langmuir frequency.
Using transformation
\ba
 \int\limits_{-\infty}^{\infty}
  e^{-\frac{m_ev_x^2}{2kT_e}}
  \frac{d v_x}
       {p_1+v_xp_2}
 &\equiv& 
 \int\limits_{0}^{\infty}
  e^{-\frac{m_ev_x^2}{2kT_e}}
  \frac{2p_1 d v_x}
       {p_1^2-v_x^2p_2^2} \simeq \nonumber\\
 &\simeq& \sqrt{\frac{2\pi kT_e}{m_e}}
            \frac{p_1}{p_1^2-\bar{v_x^2}p_2^2}\ ,
\label{eq-13}
\ea
where $\bar{v_x^2}$ can be approximated by the mean square velocity
defined by Maxwell exponent
\begin{equation}\label{eq-14}
  \bar{v_x^2} 
   \simeq
    \frac{kT_e}{m_e}\ ,
\end{equation}
one obtains
\begin{equation}\label{eq-15}
  E_{p_1p_2}
   \simeq
    \frac{p_2E_{0}/(p_1-i\omega)}
         {p_2^2 -(p_1^2/c^2)\left(1+\omega_L^2/(p_1^2-p_2^2\bar{v_x^2})\right)}
\end{equation}
and characteristic equation for the poles $p_1$, $p_2$:
\begin{equation}\label{eq-16}
  p_2^2 
  - \frac{p_1^2}{c^2}
     \left(1+\frac{\omega_L^2}{p_1^2-p_2^2\bar{v_x^2}}\right)
  = 0\ ,
\end{equation}
where it was assumed
\begin{equation}\label{eq-17}
  E(0,t) = E_0 e^{i\omega t}\ ;\qquad 
  E_{p_1}= \frac{E_0}{p_1-i\omega}\ . 
\end{equation}
This implies the pole in the complex $p_1$ plane:
\begin{equation}\label{eq-18}
  p_1 = i\omega
\end{equation}
and the corresponding pole in the complex $p_2$ plane
defined from Eq.(\ref{eq-16}).

Taking account for $\bar{v_x^2}\ll c^2$ one obtains
from Eqs.(\ref{eq-16}) and (\ref{eq-17})
two solutions:
\ba
 [p_2^{(1)}]^2
 &=&
  - \frac{\omega^2}{2\bar{v_x^2}}
     \left[
      1 + \frac{\bar{v_x^2}}{c^2}
        - \sqrt{\left(1 + \frac{\bar{v_x^2}}{c^2}\right)^2
               -4\frac{\bar{v_x^2}}{c^2}
                 \left(1-\frac{\omega_L^2}{\omega^2}\right)
               }\
     \right] \nonumber\\
\label{eq-19}
 &\simeq&
  - \frac{\omega^2}{c^2}
     \left(1 - \frac{\omega_L^2}{\omega^2}
     \right)\ ;\\
\label{eq-20}
 [p_2^{(2)}]^2
 &=&
  - \frac{\omega^2}{2\bar{v_x^2}}
     \left[
      1 + \frac{\bar{v_x^2}}{c^2}
        + \sqrt{\left(1 + \frac{\bar{v_x^2}}{c^2}\right)^2
               -4\frac{\bar{v_x^2}}{c^2}
                 \left(1-\frac{\omega_L^2}{\omega^2}\right)
               }\   
     \right]\ .
\ea  

The solution (\ref{eq-19}) at $\omega_L<\omega$,
that is
\begin{equation}\label{eq-21}
  p_2^{(1)} = ik 
  \simeq \pm i \frac{\omega}{c}
                \sqrt{1-\frac{\omega_L^2}{\omega^2}}\ ,
\end{equation}
is the well-known result of non-damping transversal
electromagnetic high-frequency waves in fully ionized plasma~\cite{bib-2}.
Phase velocity of this mode is greater than $c$,
but this result is by no means related 
to the applicability of Maxwell distribution function $f_0^{(e)}$
at high velocities $v$ up to $c$,
but, instead, is due only to the Maxwell field equation 
Eq.(\ref{eq-3}).
  
 The solution (\ref{eq-20}) for non-damping low-velocity waves
appears to be more intriguing:
\begin{equation}\label{eq-22}
  p_2^{(2)} = ik 
  \simeq \pm i \frac{\omega}{\sqrt{\bar{v_x^2}}}
                \left(1+\frac{\bar{v_x^2}\omega_L^2}{2c^2\omega^2}
                \right)
  \simeq \pm i \frac{\omega}{\sqrt{\bar{v_x^2}}}
\end{equation}
with phase and group velocities
\begin{equation}\label{eq-23}
  V_{ph} \simeq V_{gr} \simeq \sqrt{\bar{v_x^2}}\ ,
\end{equation}
which are not dependent on $\omega_L$. 

  Since we assume definiteness and convergence 
of the inverse Laplace transformation 
we should discuss an appearance of numerous poles
$(p_1+p_2v_x)=0$ in integrals in $dv_x$ in Eq.(\ref{eq-12})
depending on running values $p_1$, $p_2$.
The results of calculations of poles $p_1=i\omega$
and $p_2^{(1,2)}$, Eqs.(\ref{eq-19})--(\ref{eq-20}),
show that at these values of $p_1$ and $p_2$
the integral in $dv_x$ in Eqs.(\ref{eq-12})--(\ref{eq-13})
appears as logarithmically divergent.  
Strictly speaking,
approximation (\ref{eq-13}) implies that this integral
is defined in a principal-value sense.
In this case the inverse Laplace transformation
from approximate ``image'' to ``original'' function
is also definite both for $E_{p_1p_2}$ and $f_{p_1p_2}^{(1)}$.
The criterium for the validity of some found solutions 
$f_1(v_x,x,t)$ and $E(x,t)$ 
is the fulfilment of original Eqs.(\ref{eq-1})-(\ref{eq-3})
on these functions.
Such definite solutions can be obtained 
if one takes the principal value prescription 
for integrals in $dv_x$ (using approximation (\ref{eq-13}) or not).
This prescription is not an arbitrary agreement,
but is the necessary consequence of Laplace transform existence.
So, incorrectness  of solving Vlasov equations 
by Vlasov himself 
(in case of longitudinal plasma oscillations~\cite{bib-4})
was in representing the solution 
in the form of a single exponent $\exp(i\omega t - ikx)$
rather than in the form of $\int dv_x$ in principal value prescription.
Asymptotical solution in general case is some complex function
which can be expanded in a series of exponents,
but it is not necessarily a single exponent.

%Approximate character of expression (\ref{eq-22})
%has a principal meaning 
%since \underline{precise} value of $p_2^{(2)}$ 
%corresponds to the pole of $E_{p_1p_2}$ 
%with the main contribution to the ``original'' $E(x,t)$,
%but one should keep in mind 
%that the \underline{approximate} value
%$p_2^{(2)}\simeq \pm i\omega\sqrt{\bar{v_x^2}}$
%corresponds to the ???zero??? of $E_{p_1p_2}$ 
%with zero contribution to the ``original'' $E(x,t)$.
    Substitution of expression
\begin{equation}\label{eq-22a}
  p_{2}^{(2)}
   \simeq 
    \pm i \frac{\omega}{\sqrt{\bar{v_x^2}}}
     \left(
     1+\frac{\bar{v_x^2}\omega_L^2}{2c^2\omega^2}
     \right)
  = i k
\end{equation}
into Eq.(\ref{eq-15}) confirms the existence of two opposite in signs 
poles of the ``image" $E_{p_1p_2}$. 
Calculation of residua in these poles $p_2^{(2)}$
and amplitudes of the electric field oscillations 
for this low-velocity non-damping mode
results here in a trivial asymptotic solution 
in the form of a standing wave
$$ E(x,t)_{\mbox{asymp}} = E_0 e^{i\omega t}\cos(kx)\ . $$

However real existence and amplitudes of this mode 
must be defined 
at accounting for additive constituents 
from all other partially coupled\footnote{%
One of such additional conditions might be, for example,
the absence of a backward wave.}
or independent boundary and initial conditions (\ref{eq-11}) 
which have been omitted till now for simplicity. 
So, one of such additional conditions might be,
for example, even if partly absence of backward wave, etc.

    If the boundary and initial conditions are nevertheless 
such that some oscillatory mode with frequency $\omega$ 
is represented only with a single forward wave 
(and/or backward wave), 
all contradictions will be removed 
if the logarithmically divergent integral 
in the partial dispersion relation of this mode 
((\ref{eq-22a}) or any other) 
will be treated in the Vlasov sense of the principal value. 
In general case including collision  plasma and longitudinal waves 
one can get presence of a single travelling non-damping or damping forward wave 
only if the latter has non-exponential form 
$F(\omega t-kx)$
with a non-exponential boundary condition, 
for example, $ E = E_0 \cos(\omega t)$
(that is at least a sum of two complex-valued exponents).

    It might be interesting to note 
that quite nearly to the poles (\ref{eq-22a}) 
there are located the values of variables $p_2$ 
in the integrand $E_{p_1p_2}$ of inverse Laplace transformation
\begin{equation}\label{eq-22b}
 p_2^* = \pm i \frac{\omega}{\sqrt{\bar{v_x^2}}}  
\end{equation}
with zero contribution into Laplace $dp_2$-integral 
in the formula for an asymptotical value of $E(x,t)$. 
Such proximity between values of $p_2$ variable 
at which $E_{p_1p_2}$ turns to the infinity and to the zero is intriguing.
What will happen in case when $\bar{v_x^2}\to 0$? 
One can suppose, for example, that this case may be accompanied 
with a considerable augmentation of the length/time
needed to set up the asymptotical regime 
to really unobservable large values. 
Be that as it may, 
strict interpretation of this mode features 
is possible only at account for relativistic corrections in Eq.(\ref{eq-1}).

%%%%%%%%%%%%%%%%%%%%%%%%%%%%%%%%%%%%%%%%%%%%%%%%%%%%%%%%%%%%%%%%%%
\section{Coulomb low-collision plasma}
%%%%%%%%%%%%%%%%%%%%%%%%%%%%%%%%%%%%%%%%%%%%%%%%%%%%%%%%%%%%%%%%%%

 Coulomb collision integral is included into Eqs.(\ref{eq-1}) and (\ref{eq-12})
in the form
\ba
  f_{p_1p_2}^{(1)}
  = \frac{1}{p_1 + v_x p_2}
      \left( \frac{|e|}{m_e}\frac{\dpa f_0^{(e)}}{\dpa v_z}E_{p_1p_2}
           + Q_{p_1p_2}
      \right)\ ,\qquad\qquad\qquad&& 
  \label{eq-24}\\
  E_{p_1p_2}
      \left( p_2^2 
           - \frac{p_1^2}{c^2}
           - \frac{p_1^2}{c^2}\frac{\omega_L^2}{p_1^2-\bar{v_x^2}p_2^2}
      \right)
 = p_2 E_{p_1}
   + \frac{m_e p_1 \omega_L^2}{|e|c^2}
      \int\limits_{-\infty}^{\infty}
       \frac{v_z Q_{p_1p_2}}{p_1+v_xp_2}d\vec{v}\ ,&&
 \label{eq-25}
\ea
where $Q_{p_1p_2}$ is Laplace transform of the usual Coulomb collision integral
dominated by the electron-ion collision term~\cite{bib-2}:
\ba
 Q(\vec{v},x,t)
  &\simeq&
   \frac{2\pi e^4 L n_i}{m_e^2}\frac{\dpa }{\dpa v_i}
    \int\limits_{-\infty}^{\infty}d\vec{V}
     \frac{u^2 \delta_{ij} - u_iu_j}{u^3}
      f_0^{(i)}(\vec{V})
       \frac{\dpa f_1^{(e)}(\vec{v},x,t)}{\dpa v_j}
        \nonumber \\
  &\simeq&
   \frac{2\pi e^4 L n_i}{m_e^2}\frac{\dpa }{\dpa v_i}
    \left(
     \frac{v^2 \delta_{ij} - v_iv_j}{v^3}\
      \frac{\dpa f_1^{(e)}(\vec{v},x,t)}{\dpa v_j}
    \right)\ ,
 \label{eq-26}
\ea
where $\vec{u}=\vec{V}-\vec{v}$ and $\vec{V}$ is the ion velocity.

Thus, the calculation of the collision contribution into $E_{p_1p_2}$
reduces to the calculation of the term
\be
 \frac{p_1\omega_L^2}{c^2}
  \frac{2\pi e^4 L n_i}{m_e^2}
   E_{p_1p_2}
    \int\limits_{-\infty}^{\infty}
     \frac{v_z d\vec{v}}{p_1+v_xp_2}
      \frac{\dpa }{\dpa v_i}
       \left(
        \frac{v^2 \delta_{ij} - v_iv_j}{v^3}
        \frac{\dpa }{\dpa v_j}
        \frac{\dpa f_0^{(e)}/\dpa v_z}{p_1+v_xp_2}
       \right)\ .
 \label{eq-27}
\ee

 After integration by parts and simple transformation 
of the type (\ref{eq-13}) 
one obtains the characteristic equation for determining 
decrement $\delta$:
\ba
 \left(p_1^2-p_2^2\bar{v_x^2}\right)^4
  \left(p_2^2-\frac{p_1^2}{c^2}\right)
- \left(p_1^2-p_2^2\bar{v_x^2}\right)^3
   \frac{p_1^2\omega_L^2}{c^2} &+& \nonumber\\
+\ \frac{4\pi e^4\omega_L^2 n_i L}{3\sqrt{3\bar{v_x^2}}m_ec^2kT_e}
   p_1
    \left(p_1^6
        - 3p_1^4p_2^2\bar{v_x^2}
        + 7p_1^2\left[p_2^2\bar{v_x^2}\right]^2
        + 3\left[p_2^2\bar{v_x^2}\right]^3
    \right)
&=&0\ .\label{eq-28}
\ea

Substituting the value $p_1 = i\omega$ and assuming
\be \label{eq-29}
 p_2 = p_2^{(1)} + \delta\, \qquad 
 \left|\delta\right|
  \ll \left|p_2^{(1)}\right|\ ,       
\ee
where $p_2^{(1)}$ is defined by Eq.(\ref{eq-19}),
one obtains for the coordinate damping decrement:
\begin{equation}\label{eq-30}
  \delta \equiv \delta_1
  = \pm\frac{2\pi e^4 n_i L \omega_L^2}
            {3\sqrt{3\bar{v_x^2}} m_e kT_e c \omega^2
            \sqrt{1-\omega_L^2/\omega^2}}
\end{equation}
with $\delta_1<0$ for wave with $k<0$ (\ref{eq-21}) 
travelling to the right direction.

 But more interesting solution appears for the wave (\ref{eq-22}).
Substituting into Eq.(\ref{eq-28}) values
\be \label{eq-31}
 p_1 = i\omega\, \qquad
 p_2 = \pm i \frac{\omega}{\sqrt{\bar{v_x^2}}}
     + \delta\, \qquad 
 \left(\big|\delta\big|\ll \frac{\omega}{\sqrt{\bar{v_x^2}}}\right)\ ,       
\ee
assuming also
\be\label{eq-32}
 \big|\delta\big|
  \gg\frac{\sqrt{\bar{v_x^2}}\omega_L^2}{2\omega c^2}\ ,
\ee
and keeping in Eq.(\ref{eq-28}) only terms to the order $\delta^3$,
accounting for $\bar{v_x^2}\ll c^2$,
one obtains the following striking nonlinear result
for the low-velocity electron branch decrement:
\begin{equation}\label{eq-33}
  \delta \equiv \delta_2
  = \pm 
     \left(\frac{\pi e^4 n_i L \omega^2}
                {3\sqrt{3}\left[\bar{v_x^2}\right]^2m_e kT_e}
     \right)^{1/3}\ .
\end{equation}

  This intriguing result can be tested experimentally,
but one should have in mind 
that in low-velocity/low-frequency region 
there might also exist other electron-ion branches
defined by the ion current in the Maxwell equation (\ref{eq-3}),
which can complicate interpretation of test results.
It is worth also to note that decrements $\delta_{1,2}$,
generally speaking, should not be used automatically 
as simple additive parts of the whole collision damping decrement
of partially ionized plasma 
where collisions of electrons with neutral particles
are also present and should be taken into account.

%%%%%%%%%%%%%%%%%%%%%%%%%%%%%%%%%%%%%%%%%%%%%%%%%%%%%%%%%%%%%%%%%%
\section{Conclusions}
%%%%%%%%%%%%%%%%%%%%%%%%%%%%%%%%%%%%%%%%%%%%%%%%%%%%%%%%%%%%%%%%%%

  Application of our method~\cite{bib-1} of 2-dimensional Laplace transformation  
to plasma transversal oscillation equations 
with calculating logarithmically divergent integrals 
with the principal value prescription
results in determination of plasma oscillation frequencies (or wave numbers). 
We have obtained dispersion equations for non-damping oscillatory modes
$k(\omega)$ of quasi-neutral Maxwellian collisionless fully ionized plasma 
and damping modes of low-collision plasma 
including high-velocity (with phase velocity $>c$)
and low-velocity (with phase and group velocities $\simeq \bar{v}_e \ll c$)
transversal modes.

  With the help of the same method we have also obtained 
the damping decrements of these modes due to Coulomb electron-ion collisions
in the low-collision fully ionized plasmas.
The most striking thing is a non-linear nature of the decrement
for the low-velocity mode
$$
  \delta_2 \sim  
     \left(\frac{n_i\omega^2}{\left[\bar{v_x^2}\right]^2}\right)^{1/3}\ ,
$$  
where $\bar{v_x^2}$ is the mean-square velocity of the electrons.

 The obtained results on propagation and damping 
of plasma waves and oscillations 
can be useful not only in applications to laboratory plasma studies,
but mainly in theoretical evaluations of non-thermal energy 
and damping lengths in the solar atmosphere as well as 
in interplanetary and interstellar media.

 Let us emphasize once more: there is no necessity 
to appeal to Landau's rule of passing around poles in calculations
of indefinitely (logarithmically) divergent integrals 
both in the case of transverse as well as of longitudinal plasma waves.
The effect of dissipative ``Landau damping'', see~\cite{bib-1},
does not exist in nature 
and is no more than some abstract great fiction of theorists.
  
    It ought to rehabilitate Vlasov principal value prescription 
of his relevant logarithmically divergent integrals, 
however to generalize his solution~\cite{bib-4} 
with using not a single exponent, but some combination of exponents 
as for an asymptotical solution 
as well as for functions in boundary and initial conditions 
(for example, these conditions have to be specially selected 
 in order to avoid unphysical divergent at $x\to\infty$ backward waves, etc.).
    
    Asymptotical solution is some linear combination (a) of exponents
(Laplace expansion) which must satisfy linear plasma equations.
The linear combination (b) of the exponents from boundary and initial conditions 
ought to be considered as a selector for the exponents 
to be included in group (a).
It should be noted here that the boundary condition of the type
$ E \simeq E_0 \exp(i\omega t)$ is some mathematical abstraction 
and can not be realized in a real physical situation. 

  According to formulas given in this paper
for collisionless plasma one has
\be 
  f_{p_1p_2}^{(1)}(\vec{v})
  = \frac{1}{p_1 + v_x p_2}
      \left( \frac{|e|}{m_e}\frac{\dpa f_0^{(e)}}{\dpa v_z}E_{p_1p_2}
           + f_{p_2}^{(1)} 
           + v_x f_{p_1}^{(1)}
      \right)\,,\label{eq-36}
\ee
\be
  E_{p_1p_2}
 = \frac{1}{G}
    \left[p_2 E_{p_1}
        + F_{p_1}
        - \frac{p_1E_{p_2} + F_{p_2}}{c^2}
        - \lambda
           \int\frac{\left(p_1f_{p_1}^{(1)} 
                         - p_2f_{p_2}^{(1)}
                     \right)u_x u_z d^3\vec{u}}
                    {p_1 + u_x p_2}
    \right]
 \label{eq-37}
\ee
with $\lambda=4\pi|e|n_e/c^2$ and
\be
 G \equiv
    p_2^2 
  - \frac{p_1^2}{c^2}
  + \frac{\omega_L^2 p_1}{c^2}
     \int u_z\frac{\dpa f_0{(e)}}{\dpa u_z}
       \frac{d^3\vec{u}}{p_1 + u_x p_2}\,.
 \label{eq-38}
\ee
The point $p_1 + u_x p_2=0$ is not a point of singularity
of integrals in Eqs.(\ref{eq-37}) and (\ref{eq-38})
because the integrals are defined 
with the help of the principal value prescription.
But the point $p_1 + v_x p_2=0$
is a pole of $f_{p_1p_2}^{(1)}$ in Eq.(\ref{eq-36})
with functions $f_{p_1}^{(1)}$, $f_{p_2}^{(1)}$,
$E_{p_1}$ and others,
which, in its turn, can have singularities in $p_1$, $p_2$. 

According to~\cite{bib-5,bib-6} this pole corresponds to solutions
\be
 C(\vec{v}) e^{ik\left(x-v_xt\right)}
 \qquad 
 \left(\mbox{if}\ f_{p_2}^{(1)} \propto \frac{1}{p_2-ik}
 \right)
 \label{eq-39a}
\ee
or
\be
 C(\vec{v}) e^{i\omega\left(t-x/v_x\right)}
 \qquad 
 \left(\mbox{if}\ f_{p_1}^{(1)} \propto \frac{1}{p_1-i\omega}
 \right)
 \label{eq-39b}
\ee
of equation
\begin{equation}
 \label{eq-40}
  \frac{\dpa f_1^{(e)}}{\dpa t}
  + v_x \frac{\dpa f_1^{(e)}}{\dpa x}
  = 0
\end{equation}
with arbitrary $k$ (or $\omega$), $C(\vec{v})$, and $v_x$.

Inverse Laplace transform $\varphi(\vec{v},x,t)$
of expression
\be
 \varphi_{p_1p_2} =
 \frac{\varphi_{p_1}(\vec{v})+v_x\varphi_{p_2}(\vec{v})}
      {p_1 + v_x p_2}\,,
 \label{eq-41}
\ee
with some arbitrary initial ($\varphi_{p_2}$) 
and boundary ($\varphi_{p_1}$) conditions
gives the general complete solution of Eq.(\ref{eq-40}),
including as well 
the alternative partial solutions (\ref{eq-39a}) and (\ref{eq-39b}).
So, one can always select from the set of solutions
of Eq.(\ref{eq-1})
\begin{equation}
 \label{eq-42}
  f_1^{(e)}(\vec{v},x,t;\varphi) = 
  f_1^{(e)}(\vec{v},x,t) +
  \varphi(\vec{v},x,t)
\end{equation}
with arbitrary initial and boundary conditions
a variant with 
$f_{p_1}^{(e)}=f_{p_2}^{(e)}=0$.
This solution, due to Eq.(\ref{eq-37}),
induces appearance of electrical field $E(x,t)$
and of unphysical backward waves.

Let us assume
the boundary electric field to be of the following form
\begin{equation}
 \label{eq-43}
  E(0,t) = 
  E_0 \cos(\omega t)\,.
\end{equation}
Then it is natural to expect
\begin{equation}
 \label{eq-44}
  f_1^{(e)}(\vec{v},0,t) = 
  a_\omega(\vec{v})\cos(\omega t) +
  b_\omega(\vec{v})\sin(\omega t)\,
\end{equation}
with some arbitrary coefficients 
$a_\omega(\vec{v})$ and $b_\omega(\vec{v})$.

In this case Laplace transforms are
\ba
 E_{p_1} &=&
 \frac{p_1E_0}{p_1^2+\omega^2}
 \,;\label{eq-45}\\
 f_{p_1}^{(1)} &=&
 \frac{p_1a_{p_1}(\vec{v})+|p_1|b_{p_1}(\vec{v})}{p_1^2+\omega^2}
 \,.\label{eq-46}
\ea

    According to Eq.(\ref{eq-37}) with account for collision
damping 
(through the additional collision term in $G$)
the term $p_2E_{p_1}/G$ leads to asymptotical waves with exponents of 
$(i\omega t \pm ikx \pm \delta x)$ 
and
$(-i\omega t \pm ikx \mp \delta x)$,
and the term with $f_{p_1}^{(1)}$ leads to waves with exponents of 
$(i\omega t \pm ikx \pm \delta x)$ (with coefficient $a_{\omega}$ 
in Eq.(\ref{eq-46})) and
$(-i\omega t \pm ikx \mp \delta x)$ (with coefficient $b_{\omega}$
in Eq.(\ref{eq-46})). 
It is evident that it is always possible 
to eliminate backward waves 
by the proper selection of $a_{\omega}$ and $b_{\omega}$. 
Possible more convenient modification of Eq.(\ref{eq-46})
can be also, for instance, 
\begin{equation}\label{eq-47}
 f_{p_1}^{(1)} =
 \frac{p_1a_{p_1}(\vec{v})+p_1^2b_{p_1}(\vec{v})}
      {p_1^2+\omega^2}
      \,,
\end{equation} 
where $a_{p_1}(\vec{v})$, $b_{p_1}(\vec{v})$, and $f_{p_1}^{(1)}$
are all proportional to $E_0$.

    At account for the simultaneous presence
of the other mode (\ref{eq-22}),
it is seen that the already defined above 
$a$ and $b$ 
have to lead to removal of backward waves in this mode too,
due to its analogous structure.

    The paradox consists in the fact that 
at damping electrical field\footnote{%
For instance, in low-collision plasma.},
due to collision terms in $G$,
there exist non-damping "ballistic" kinematical oscillations 
in $f_{1}^{(e)}(\vec{v},x,t)$
(in ~\cite{bib-5,bib-6} there was implied, instead,
 collisionless Landau damping with the same result). 
These "ballistic" oscillations appear to be unremovable. 
But there is no mystery. 
If collision term, even if partly, 
included some additional term of the form
\be
 - C(\vec{v}) 
 f_{1}^{(e)}(\vec{v},x,t)
 \,,\label{eq-48}
\ee  
then, instead of denominator $p_1+v_xp_2$ 
and $p_1+u_xp_2$ in Eqs.(~\ref{eq-36})-(~\ref{eq-38}),
one would have denominator 
$p_1+v_xp_2+C(\vec{v})$
or $p_1+u_xp_2+C(\vec{u})$
and the "ballistic" term would be damping too
in the same extent as the oscillations of electric field.  

    It appears that, for instance, as the first collision iteration  
one could more reasonably use 
the substitution in Eq.(\ref{eq-24})
\be
 Q_{p_1p_2}
  \equiv
   f_{p_1p_2}^{(1)}
    \left(\frac{Q_{p_1p_2}}{f_{p_1p_2}^{(1)}}\right)
   \simeq
   f_{p_1p_2}^{(1)}
    \left[\frac{Q_{p_1p_2}^{o}}
               {\left(f_{p_1p_2}^{(1)}\right)_o}
    \right]
   \,,
 \label{eq-49}
\ee
where $\left(f_{p_1p_2}^{(1)}\right)_o$ is 
the collisionless approximation of $f_{p_1p_2}^{(1)}$ 
according to Eq.(\ref{eq-36}),
and $Q_{p_1p_2}^{o}$ is the function of $p_1$, $p_2$, and $\vec{v}$
defined by the Laplace transformation 
of expression (\ref{eq-26})
with substitution 
$f_{p_1p_2}^{(1)}$ $\to$ $\left(f_{p_1p_2}^{(1)}\right)_o$. 
Here arises also the need for some new correction of coefficients 
$a$ and $b$ in the boundary and initial conditions 
$f_{p_1}^{(e)}$ ad $f_{p_2}^{(e)}$
in order to avoid growing backwards waves.

    By this way non-damping of ballistic mode 
in low-collision plasma can be related with some approximation
made in iteration procedures 
(yet nobody has still investigated convergence 
of the related iteration processes)
and not with intrinsic features of the Coulomb collision integral.
 
    This example of a successful application of our methodology
to plasma oscillations supports the hope of further unraveling these
and more entangled problems of plasma-echo phenomena.
In particular, in the case of boundary conditions 
with two superimposed frequencies 
\begin{equation}
 \label{eq-51}
  E(0,t) = 
  E_0 \cos(\omega_1t)
       \cos(\omega_2t)
\end{equation}
two non-damping waves 
with frequencies $(\omega_1-\omega_2)$ and
$(\omega_1+\omega_2)$
arise in collisionless plasma.
That is a very simple imitation of plasma echo effect
(cf. description in~\cite{bib-6})
for both transversal and longitudinal oscillations.
\vspace{5mm}

\textbf{Acknowledgements}
  The author is thankful to Dr.~A.~P.~Bakulev 
for his criticism and assistance in preparing the paper in \LaTeX\ style,
without which this paper never would be completed.

\end{document}